\documentclass[prd,aps,floats,preprint,preprintnumbers,nofootinbib]{revtex4}
\usepackage{amsmath,amssymb,mathrsfs}
\usepackage{graphicx}
\usepackage{dsfont}
\usepackage{graphics}
\usepackage{amsmath,amscd}
\usepackage{epsfig}
\usepackage{latexsym,oldgerm}
\def\beq{\begin{equation}}
\def\eeq{\end{equation}}
\def\bea{\begin{eqnarray}}
\def\eea{\end{eqnarray}}

\newcommand{\G}{\mathscr{G}}
\newcommand{\Po}{\mathscr{P}}
\newcommand{\Pa}{\mathbb{C}\mathscr{P}}
\newcommand{\Pat}{\mathbb{C}\mathscr{P}_{\theta}}
\newcommand{\Ga}{\mathbb{C}\G}
\newcommand{\Tw}{\mathcal{F}_{\theta}}

\newcommand{\e}{{\rm e}}
\newcommand{\state}{\langle a',\Lambda'|}
\newcommand{\Fstate}{\langle\e_p\otimes\alpha|}
\newcommand{\dx}{{\rm d}}
\newcommand{\mut}{\mu_{\theta}}
\newcommand{\muts}{\tilde{\mu}_{\theta}}
\newcommand{\I}{\mathds{1}}

\begin{document}
\small
\preprint{SU-4252-884 \vspace{1cm}} \setlength{\unitlength}{1mm}
\title{Space-time from Symmetry: The Moyal Plane from the Poincar\'e-Hopf Algebra  
\vspace{0.5cm}}
\author{ A. P.
Balachandran$^{a,b,\ddagger}$}\thanks{C\'atedra de Excelencia\\$\ddagger$bal@phy.syr.edu} \author{M. Martone$^{a,c}$}\thanks{mcmarton@syr.edu}
\affiliation{$^{a}$Department of Physics, Syracuse University, Syracuse, NY
13244-1130, USA\\
$^{b}$Departamento de Matem\'aticas, Universidad Carlos III de Madrid, 28911 Legan\'es, 
Madrid, Spain\\
$^{c}$Dipartimento di Scienze Fisiche, University of Napoli and INFN, Via Cinthia I-80126 Napoli, Italy}
\begin{abstract}
\vspace{0.5cm} 
We show how to get a non-commutative product for functions on space-time starting from the deformation of the coproduct of the Poincar\'e group using the Drinfel'd twist. Thus it is easy to see that the commutative algebra of functions on space-time ($\mathbb{R}^4$) can be identified as the set of functions on the Poincar\'e group invariant under the right action of the Lorentz group provided we use the standard coproduct for the Poincar\'e group. We obtain our results for the noncommutative Moyal plane by generalizing this result to the case of the twisted coproduct. This extension is not trivial and involves cohomological features.

As is known, spacetime algebra fixes the coproduct on the dffeomorphism group of the manifold. We now see that the influence is reciprocal: they are strongly tied.

\end{abstract}
\maketitle
\section{INTRODUCTION}\label{sec:intro}

During the last fifteen years, excellent physical arguments have emerged suggesting that spacetime at Planck scale is noncommutative. In particular it has been argued \cite{Doplicher} that the coexistence of Einstein's theory of relativity and basic quantum mechanics, namely Heisenberg's uncertainty principle, makes the quantum nature of spacetime important at energy scales close to Planck scale. A simple model which reflects this noncommutativity is the Moyal plane $\mathcal{A}_{\theta}(\mathbb{R}^N)$. It is defined by the $*$-product
\beq\label{Md21}
f_1*f_2(x)=f_1(x)\textrm{e}^{\frac{i}{2}\theta_{\alpha \beta}\overleftarrow{\partial^{\alpha}}\otimes \overrightarrow{\partial^{\beta}}} f_2(x)
\eeq
on functions $f_i$ on $\mathbb{R}^N$. The latter implies the commutation relation
\beq \label{Md1}
[\widehat{x}_{\mu}, \widehat{x}_{\nu}] = i \theta_{\mu \nu} 
\eeq
where $\theta_{\mu \nu} = - \theta_{\nu \mu}$ are constants and
$\widehat{x}_{\mu}$ on the coordinate functions: 
\beq
\widehat{x}_{\mu}(x) = x_{\mu}. 
\eeq

Now the noncommutative multiplication rule (\ref{Md21}) can be written using the so called {\it Drinfel'd twist} $\mathcal{F}_{\theta}$ \cite{drinfeld}:
\bea\label{Md7}
f_1*f_2&=&m_{\theta}(f_1\otimes f_2)=m_0\circ\mathcal{F}_{\theta}(f_1\otimes f_2)\\
\mathcal{F}_{\theta}&:=&\exp{\left(\frac{i}{2}\theta_{\mu\nu}\partial^{\mu}\otimes\partial^{\nu}\right)}
\eea
where $m_0$ is the commutative, point-wise multiplication and $m_{\theta}$ is the noncommutative one given by the noncommutativity in the coordinates (\ref{Md1}).

Also it was thought for a long time that the relation (\ref{Md1}) spoils Poincar\'e invariance completely since under the naive Lorentz transformations of $\hat{x}_{\mu}$ and $\hat{x}_{\nu}$, the L.H.S. of (\ref{Md1}),   transforms in a non-trivial way whereas the R.H.S. being a constant does not change. Later, the Poincar\'e symmetry was restored (at least partially \cite{wess,chaichian, Aschieri1, Aschieri2}) by changing the action of the Poincar\'e group to the so-called twisted action \cite{sasha}. It involves a deformation of the coproduct $\Delta_0$ of the Poincar\'e-Hopf algebra to a coproduct $\Delta_{\theta}$ for compatibility with the product (\ref{Md7}) (for a discussion on Hopf algebras see \cite{aschieri, chari, majid}). The deformation is as follows:
\bea\label{Md8}
\Delta_0&\to&\Delta_{\theta}=F_{\theta}^{-1}\Delta_0F_{\theta},\\\label{Md10}
F_{\theta}&=&\exp{\left(-\frac{i}{2}\theta_{\mu\nu}P^{\mu}\otimes P^{\nu}\right)}\\
P^{\mu}&=&Translations\ generators
\eea
(Note that $\mathcal{F}_{\theta}$ is the realization of $F_{\theta}$ on functions on $\mathbb{R}^N$.)

In this paper, we will show that from the deformation (\ref{Md8}) of the Hopf algebra, it is possible to get a deformation on the algebra of functions on the Poincar\'e group and then obtain the Moyal algebra $\mathcal{A}_{\theta}(\mathbb{R}^N)$ therefrom using non-trivial considerations. This is done by constructing the dual Hopf algebra to the Poincar\'e-Hopf algebra and using the fact that $\mathbb{R}^N$ can be identified with the Poincar\'e group quotiented by the Lorentz group. For the standard coproduct ($\theta_{\mu\nu}=0$), the procedure leads to the commutative algebra of functions on spacetime. The dual Hopf algebra and its relevant properties are recalled in section 2. Subsequent sections discuss the construction of spacetime algebra therefrom.

\section{Hopf algebra duality}

We will briefly recall now how to construct the dual of a Hopf algebra, for details see \cite{chari,majid}. 
Given a Hopf algebra $(H,\Delta,\mu,\eta,\epsilon,\mathcal{S})$, where $\Delta$ is the co-product, $\mu$ is the multiplication map, and $\eta$, $\epsilon$ and $\mathcal{S}$ are respectively the unit, co-unit and the antipode, we want to construct another Hopf algebra $(H^*,\Delta^*,\mu^*,\eta^*,\epsilon^*,\mathcal{S}^*)$ which will be called the {\it dual of H}.  Since we do not want to deeply go into mathematical formality, we will just focus on how to get $\Delta^*$ and $\mu^*$ from $H$ assuming that once we find these structures, the unit, co-unit and the antipode can also be found.

Now $H^*$ is the dual of $H$. We want to identify $H^*$ with (linear) functions $\mathscr{F}(H)$ on $H$. For this, we need a pairing of functions $\in\mathscr{F}(H)$ with elements of $H$. For this, we use the natural choice:
\beq
\forall f\in \mathscr{F}(H),\ h\in H\quad \langle f,h\rangle\equiv f(h)\in \mathbb{C}\quad.
\eeq
With this pairing, we identify $H^*$ with $\mathscr{F}(H)$. Then we can define $\mu^*$ as follows:
\beq\label{Md2}
\mu^*:\forall f_1,\ f_2 \in H^*, h \in H \quad \langle \mu^*(f_1\otimes f_2), h\rangle :=\langle f_1\otimes f_2, \Delta(h)\rangle\ {\rm or} \ (f_1\cdot^*f_2)(h):=(f_1\otimes f_2)(\Delta(h))
\eeq
where we have indicated $\mu^*(f_1\otimes f_2)$ as $f_1\cdot^*f_2$.

In the same way, we can construct the $coproduct$ on $H^*$:
\beq\label{Md3}
\Delta^*:\forall h_1,h_2\in H, f\in H^*\quad \langle \Delta^*(f),h_1\otimes h_2\rangle:=\langle f,\mu(h_1\otimes h_2)\rangle \ {\rm or}\ \Delta(f)(h_1\otimes h_2):=f(h_1\cdot h_2)
\eeq
where in this case we have indicated $\mu(h_1\otimes h_2)$ as $h_1\cdot h_2$. From the relations above, it is clear that the {\it co-structure} of $H$ will give rise to the multiplication in $H^*$ and vice-versa. In particular if we deform, in the sense of the Hopf algebra deformation theory \cite{chari,majid}, the co-product of the first one, it will translate into a deformation of the multiplication rule for $H^*$. This already suggests that somehow starting from the co-deformation of the Poincar\'e-Hopf algebra, namely the Poincar\'e group algebra (see below) using a deformed co-product, we can get a deformation of the multiplication of the dual, that is of functions on the Poincar\'e group, which can then be translated into a deformation of the algebra of functions on spacetime.

Once the pairing is given, there is also an intrinsic and natural way of lifting the left and right actions of the group on itself,
\bea
&&\rho_L(\tilde{h})\triangleright h:=\tilde{h}\cdot h\\
&&\rho_R(\tilde{h})\triangleright h:=h\cdot \tilde{h}^{-1}
\eea
to the dual $H^*$. It goes as follows:
\bea\label{Md5}
&&f_L\equiv\rho^*(\tilde{h})\triangleright_L f\quad:\quad \langle f_L,h\rangle:=\langle f,\rho_L(\tilde{h}^{-1})\triangleright h\rangle=f(\tilde{h}^{-1}\cdot h)\\\label{Md6}
&&\ f_R\equiv\rho^*(\tilde{h})\triangleright_R f\quad:\quad \langle f_R,h\rangle:=\langle f,\rho_R(\tilde{h}^{-1})\triangleright h\rangle=f(h\cdot\tilde{h})
\eea
The action on $(H\otimes H)^*$ is lifted, similarly, using the co-product $\Delta$. 

These observations will be very useful for us in the following.

As a final remark, we emphasise that to identify $(H\otimes H)^*$ with say $\mathscr{F}(H)\otimes\mathscr{F}(H)$, we need a pairing of the latter with $H\otimes H$. In other words, we need an identification of $H^*\otimes H^*$ with $(H\otimes H)^*$ by introducing a pairing of the former with $H\otimes H$. A suitable pairing has been assumed in (\ref{Md3}). An example of this ambiguity is already present in the possibility of taking the ``flipped pairing'', $\langle f_1\otimes f_2|g_1\otimes g_2\rangle=f_1(g_2)f_2(g_1)$. The significance of this remark will emerge below.

\section{The Group Algebra and its dual}

To get a feeling of the construction above as well as to introduce the two Hopf algebras we will be working with, we want now to construct the dual of a specific Hopf algebra, namely the group algebra $\Ga$ of a group $\G$, regarded as a Hopf algebra, in terms of functions on $\Ga$.

For a given group $\G$, its group algebra is the vector space over complex numbers obtained from any linear combination of the elements of the group upon which we define a multiplication and a co-multiplication rule. The product between two elements $g_1=\sum_k\lambda_k\cdot g_k$ and $g_2=\sum_l\theta_l\cdot g_l$ of $\Ga$ where $ \lambda_k,\theta_l\in{\mathbb{C}}$ and $g_k,g_l\in\G$ is inherited from group multiplication:
\beq
g_1\cdot_{\Ga} g_2:=\sum_{k,l}(\lambda_k\theta_l)g_k\cdot_{\G} g_l
\eeq 
where $\cdot_{\G}$ is the group multiplication of $\G$. The canonical co-product $\Delta$ is defined as follows:
\beq\label{Md4}
\forall g\in\G, \Delta(g)=g\otimes g
\eeq
Using linearity we can extend $\Delta$ to the whole group algebra.

We want now to find its dual, namely the Hopf algebra of functions upon the group $\G$. We identify functions $f$ on $\G$ with elements of $\Ga^*$ using the natural pairing
\beq\label{Md9}
\langle f,g\rangle=f(g),\quad g\in\G
\eeq
where $f(g)$ is just the value of the function on the point $g$. Again we can extend the pairing of $f$ to all elements of $\Ga$ using linearity. 

It is now time to use (\ref{Md2}-\ref{Md3}) to get the structures induced by the ones on $\Ga$.

First let us identify $\Ga^*\otimes\Ga^*$ with $(\Ga\otimes\Ga)^*$ by assuming the pairing
\beq
\langle f_1\otimes f_2,g\otimes g\rangle=f_1(g)f_2(g),\quad f_i\in\Ga^*,\quad g\in\G.
\eeq
This pairing is then as usual extended to all of $\Ga\otimes\Ga$ using linearity.

The multiplication rule for functions on $\G$ is, using (\ref{Md4}):
\beq\label{Md20}
\langle\mu^*(f_1\otimes f_2),g\rangle=\langle f_1\otimes f_2,g\otimes g\rangle\ {\rm or}\ (f_1\cdot^* f_2)(g)=f_1(g)f_2(g)
\eeq
which is simply the point-wise, commutative, product. The co-product induced is instead less trivial:
\beq\label{Md23}
\langle\Delta^*(f),g_1\otimes g_2\rangle=\langle f,g_1\cdot_{\G} g_2\rangle\ {\rm or}\ \Delta^*(f)(g_1\otimes g_2)=f(g_1\cdot_{\G}g_2) 
\eeq
We thus find that the Hopf algebra dual to $\Ga$ is nothing but the commutative algebra of functions on the group $\G$, with the co-product above. 

We emphasise that to arrive at (\ref{Md20}), we have chosen a specific pairing between $f_1\otimes f_2$ and $g\otimes g$. We can instead choose another pairing $\langle f_1\otimes f_2,g\otimes g\rangle'=\langle f_1\otimes f_2,K(g\otimes g)\rangle$ where $K$ is an invertible linear operator chosen from $\Ga\otimes\Ga$ without setting $K=\I\otimes\I$. This would have different consequences.

The actions given by (\ref{Md5}-\ref{Md6}) in this case give the {\it right} and {\it left} regular representations of the group $\G$ on the functions. If we choose the Poincar\'e group $\Po$ for $\G$, the algebra of functions on spacetime can be obtained as the coset of the dual of $\Pa$ with respect to the right action of the Lorentz group. This can be understood since spacetime is topologically $\mathbb{R}^4$ and so is the coset $\mathscr{P}\diagup\mathscr{L}^{\uparrow}_+$. We notice that in this case the construction can be consistently carried through since the point-wise product of right-invariant functions under the Lorentz group is still right invariant. This will not be the case when we consider the dual of the deformed Poincar\'e group algebra, this problem has been already investigated, look at \cite{Jerzy} and references therein\footnote{In these approaches the quantum spacetime algebra is provided by the translation sector of the dual of the deformed Hopf-Poincar\'e. In this case we end up with different commutation relation among spacetime coordinates in which an ``anomalous'' term dependent upon Lorentz group parameter appears.}. We will now show a possible way of circumventing this issue using the above duality construction.

\section{Moyal from Poincar\'e}

In this section we show how to get Moyal spacetime applying the above considerations to the deformation $\Pat$ of the Poincar\'e group algebra. $\Pat$ is a particular Hopf algebra deformation of $\Pa\equiv\Pa_0$ in which only the co-product has been deformed (see (\ref{Md8})). Since the co-product in $\Pat$ is no longer co-commutative, we expect the multiplication induced on the algebra of functions by duality to be noncommutative like the Moyal one in fact.

As we said at the end of the previous section, the problem in this case is that the preceding coset operation cannot be carried through in an obvious manner. In order to be able to consider right-invariant functions on $\Pat$, we have to modify the pairing of $f_1\otimes f_2$ with $\Pat\otimes\Pat$ where $f_i$ are functions on $\Pat$ with the natural pairing (\ref{Md9}) from which we start the construction. This modification of the pairing, gotten by just asking for compatibility with invariance under the right action of the Lorentz group, turns out to be exactly what we need to obtain the multiplication rule (\ref{Md7}) on the algebra of function on $\mathbb{R}^4$. As mentioned above we will present the calculation in the example of the Moyal twist, $F_{\theta}=\exp\left(-\frac{i}{2}P^{\mu}\theta_{\mu\nu}\otimes P^{\nu}\right)$, that is enough to see the procedure for say the Wick-Voros twist. 

$\Pat$ can be thought of as generated by elements $U(a,\Lambda)$, where $U(a,\Lambda)=U(a)U(\Lambda)$ is a faithful realization of $\Po$. Also:
\beq\label{Md13}
U(\Lambda)U(a)=U(\Lambda a)U(\Lambda).
\eeq 
Using (\ref{Md13}), we can write all the elements of $\Pat$ with the translation on the left and Lorentz transformation on the right. This is hereafter assumed. 

Note also that
\beq\label{Md11}
(b\cdot P)U(a)=-i\frac{d}{d\lambda}U(a+\lambda b)\Big|_{\lambda=0}
\eeq
where $P_{\mu}$ is the generator of translations. 

It is convenient now to find a basis for the dual. Following Dirac, any element in $\Pat^*$ can be written in terms of the $\delta$-distributions $\state$ where
\beq
\state U(a)U(\Lambda)\rangle=\delta^4(a'-a)\delta(\Lambda'^{-1}\Lambda)
\eeq

Fourier transforming in $a'$ we can thus regard $\Pat^*$ as spanned by
\beq
\Fstate\quad:\quad \Fstate U(a)U(\Lambda)\rangle=\int\dx^4a' \e^{ip'\cdot a'}\langle a',\alpha| U(a)U(\Lambda)\rangle=\e_p(a)\alpha(\Lambda)
\eeq
where $\e_{p}(a)=\e^{ip\cdot a}$ and $\alpha(\Lambda)\in\mathbb{C}$. Note that:
\beq
\Fstate U(\Lambda)U(a)\rangle=\Fstate U(\Lambda a)U(\Lambda)\rangle=\e^{ip\cdot\Lambda a}\alpha(\Lambda).
\eeq

We are now ready to compute explicitly the deformed product $\mu^*$ induced on $\Pat^*$. Recalling (\ref{Md8}), (\ref{Md10}) and (\ref{Md2}) and with $f_1=\e_p\otimes \alpha$ and $f_2=\e_q\otimes \beta$,
\bea
\langle \mut^*(f_1\otimes f_2)|U(a)U(\Lambda)\rangle&=&\langle f_1\otimes f_2|\Delta_{\theta}(U(a)U(\Lambda))\rangle=\langle f_1\otimes f_2|F^{-1}_{\theta}\Big[U(a)U(\Lambda)\otimes U(a)U(\Lambda)\Big]F_{\theta}\rangle\nonumber\\
&=&\langle\e_{p}\alpha\otimes\e_{q}\beta|\e^{\frac{i}{2}P\wedge P-\frac{i}{2}(\Lambda P)\wedge(\Lambda P)}\Big[U(a)U(\Lambda)\otimes U(a)U(\Lambda)\Big]\rangle
\eea
where $P\wedge P=P^{\mu}\theta_{\mu\nu}\otimes P^{\nu}$. To evaluate the expression above, we need to know how to evaluate expressions like
\beq
\langle \e_p|(b\cdot P)U(a)\rangle
\eeq
From (\ref{Md11}), this is given by
\beq
-i\frac{d}{d\lambda}\langle \e_p|U(a+\lambda b)\rangle\Big|_{\lambda=0}=b\cdot p\ \e^{ip\cdot a}
\eeq
where $P$ has become $p$. Then pairing $f_1\otimes f_2$ with $\Pat\otimes\Pat$,
\beq\label{Md12}
\langle\mu_{\theta}^*(f_1\otimes f_2)|U(a)U(\Lambda)\rangle:=(f_1*f_2)\Big(U(a)U(\Lambda)\Big)=\e^{-\frac{i}{2}(p\wedge q)+\frac{i}{2}(\Lambda p)\wedge(\Lambda q)}\ \e_{p+q}(a)\alpha(\Lambda)\beta(\Lambda)\quad.
\eeq
where in the evaluation, following \cite{sasha}, we have used what above has been called ``flipped pairing''. This pairing is assumed from now on for convenience since otherwise finally we will end up with $\mathcal{A}_{-\theta}(\mathbb{R}^N)$.

Using the right and left action on $\Pat^*$ described above (\ref{Md5}-\ref{Md6}), this can be written as
\beq\label{Md18}
\mut\circ(\e_p\alpha\otimes\e_q\beta)=\mu_0\Big[\e^{-\frac{i}{2}P^L\wedge P^L-\frac{i}{2}P^R\wedge P^R}\e_p\alpha\otimes\e_q\beta\Big].
\eeq

Before taking the coset with respect the right action of the Lorentz group, we need to ensure that it can be taken consistently. This is the case if the product of two right-invariant functions is still right-invariant. From the explicit expression (\ref{Md12}), we can already see that the product acquires a dependence on $\Lambda$ spoiling this condition. More formally, in the basis we have chosen, a right invariant function on $\Pat^*$ is one in which the ``Lorentz part'' $\alpha$ is the constant function. Now the Lorentz group does not act on the right trivially on the product of two such functions. Thus from (\ref{Md6}):
\bea
&&\rho_R^*\Big(U(a')U(\Lambda')\Big)\triangleright(f_1*f_2)\Big(U(a)U(\Lambda)\Big)=\langle \e_p\alpha\otimes\e_q\beta|\Delta_{\theta}\Big(U(a+\Lambda a')U(\Lambda\Lambda')\Big)\rangle\nonumber\\\label{Md14}
&&\qquad\qquad=\e^{-\frac{i}{2}p\wedge q}\e^{\frac{i}{2}(\Lambda\Lambda'p)\wedge(\Lambda\Lambda'q)}\e_{p+q}(a+\Lambda a')\alpha(\Lambda\Lambda')\beta(\Lambda\Lambda')
\eea
where we used the compatibility of the co-product with the product multiplication in $\Pat$, \newline$\Delta_{\theta}(g)\cdot\Delta_{\theta}(g')=\Delta_{\theta}(g\cdot g')$ and (\ref{Md13}). The equation above shows that there is a non-trivial dependence on the Lorentz group coming from twisting the co-product. That prevents us from taking the coset and still retaining an algebra structure. Thus from the product (\ref{Md12}) obtained from the pairing (\ref{Md9}), we cannot proceed with the coset and obtain an algebra of functions on the spacetime.

\section{Moyal from a modified pairing}

As we already emphasised above, in the identification of $f_1\otimes f_2$ ($f_i\in\Pat^*$) with an element of ($\Pat\otimes\Pat$)$^*$ we have a certain freedom. Thus we can carry out the whole construction modifying how $\Pat^*\otimes\Pat^*$ pairs with $\Pat\otimes\Pat$. If $\sigma$ be an invertible element of $\Pat\otimes\Pat$, then a possible pairing is
\beq\label{Md16}
\langle f_1\otimes f_2|g_1\otimes g_2\rangle_{\sigma^R}=(f_2\otimes f_1)\Big((g_1\otimes g_2)\circ\sigma\Big)\equiv(f_2\otimes f_1)\Big(\sigma^{R}\circ(g_1\otimes g_2)\Big)
\eeq
where composition $\circ$ should be understood in terms of products and actions on the Hopf algebra.  In the previous construction we implicitly assumed the ``trivial'' pairing (with $\sigma=\I\otimes\I$) of $f_1\otimes f_2$ with elements of $\Pat\otimes\Pat$. 

The modification (\ref{Md16}) does not effect the induced co-product $\Delta^*_{\theta}$ whereas the new pairing defines a new multiplication map $\muts^*$\footnote{We are not considering here compatibility of the two structures ($\Delta^*_{\theta}$,$\muts^*$). This will be briefly discussed below.}:
\beq\label{Md17}
\langle\muts^*(f_1\otimes f_2)|h\rangle_{\sigma^R}=\langle f_1\otimes f_2|\Delta_{\theta}(h)\rangle_{\sigma^R}=(f_2\otimes f_1)\Big(\sigma^R\circ\Delta_{\theta}(h)\Big).
\eeq

A natural question to pose is under what conditions $\sigma$ gives rise to an associative product. We will show in the final section that associativity of $\muts^*$ puts cohomological constraints on $\sigma$ in the sense of the Hopf algebra deformation theory \cite{majid}.

With a particular choice of the map $\sigma$, we can ensure that the product (\ref{Md12}) is compatible with the right invariance under the action of the Lorentz group. Thus if we assume that
\beq\label{Md22}
\sigma\equiv\sigma^R=\e^{\frac{i}{2}P^R\wedge P^R}
\eeq
and use (\ref{Md17}) and the expression (\ref{Md18})  for $\mut^*$, we get an expression for a new multiplication map:
\beq\label{Md19}
\muts^*(\e_p\alpha\otimes\e_q\beta)\Big(U(a)U(\Lambda)\Big)=\Big[\mu_0\Big(\e^{-\frac{i}{2}P^L\wedge P^L}\e_p\alpha\otimes\e_q\beta\Big)\Big]\Big(U(a)U(\Lambda)\Big)=\Big[\mu_0\Tw(\e_p\alpha\otimes\e_q\beta)\Big]\Big(U(a)U(\Lambda)\Big)
\eeq
where $\Tw$ acts only on $\e_p$ and $\e_q$.

Acting on the right, we can check that the choice of pairing made above allows us to take cosets. Thus now
\beq
\rho_R^*\Big(U(a')U(\Lambda')\Big)\triangleright\muts^*(f_1\otimes f_2)\Big(U(a)U(\Lambda)\Big)=\e^{-\frac{i}{2}p\wedge q}\e_{p+q}(a+\Lambda a')\alpha(\Lambda\Lambda')\beta(\Lambda\Lambda')
\eeq
Assuming that $\alpha$ and $\beta$ are invariant under the right action of the Lorentz group (that is that they are constant functions on the Lorentz group) and putting   $a'=0$, we get from above that also the product is right-invariant under Lorentz transformations. This is what we required.

After the coset $\Pat^*\diagup\mathscr{L}^{\uparrow}_+ $ has been taken, we thus obtain a deformation of the algebra of functions on spacetime with product $\muts^*$. A comparison of (\ref{Md19}) with (\ref{Md7}) shows that this deformation of the algebra of functions is exactly the one given by Moyal twist $m_{\theta}=m_0\circ\Tw$, just as we claimed.

\section{Rigidity of Spacetime algebra}

The above construction only involves a modification of the pairing on the right. Asking for the multiplication map in (\ref{Md17}) to preserve the invariance of functions under the right-action of the Lorentz group does not however give rise to constraints on possible modifications of the pairing on the left. Thus the following choice would still be compatible with the above construction:
\beq
\langle f_1\otimes f_2|g_1\otimes g_2\rangle_{\delta^L\circ\sigma^R}:=(f_2\otimes f_1)\Big((\delta^L\circ\sigma^R)\circ(g_1\otimes g_2)\Big)\equiv(f_2\otimes f_1)\Big(\delta\circ(g_1\otimes g_2)\circ\sigma\Big)
\eeq
where $\delta$ is an invertible element of $H\otimes H$. This coupling would define a new product, namely:
\beq\label{Md27}
\muts'^*(f_1\otimes f_2)(g):=(f_2\otimes f_1)(\delta\circ\Delta_{\theta}(g)\circ\sigma)
\eeq

We will show that this freedom in changing the pairing from the left can be ruled out by requiring that there is a left action of the Hopf-Poincar\'e group on the algebra of functions we get from the coset operation. Thus we require that\beq
\rho^*(\tilde{h})\ \triangleright_L\Big(\muts'^*(f_1\otimes f_2)\Big)(h)=\muts'^*\Big(\rho_{H^*\otimes H^*}^*(\tilde{h})\ \triangleright_L(f_1\otimes f_2)\Big)(h)
\eeq
where $\rho_{H^*\otimes H^*}^*(\tilde{h})=\Big(\rho^*_{H^*}\otimes\rho^*_{H^*}\Big)\Delta(\tilde{h})$ and $\rho_{H^*}^*$ is defined by (\ref{Md5}). We can now compute both sides:
\beq\label{Md25}
{\rm L.H.S.}=\muts'^*(f_1\otimes f_2)(\tilde{h}^{-1}\cdot h)=(f_2\otimes f_1)\Big(\delta\circ\Delta(\tilde{h}^{-1})\circ\Delta(h)\circ\sigma\Big)
\eeq
where compatibility between $\Delta_{\theta}$ and group composition law has been used. 

On the R.H.S. we get:
\beq\label{Md26}
{\rm R.H.S.}=\Big(\rho_{H^*\otimes H^*}^*(\tilde{h})\ \triangleright_L(f_2\otimes f_1)\Big)(\delta\circ\Delta(h)\circ\sigma)=(f_2\otimes f_1)\Big(\Delta(\tilde{h}^{-1})\circ\delta\circ\Delta(h)\circ\sigma\Big)
\eeq

Since (\ref{Md25}) and (\ref{Md26}) have to be equal for all elements $\tilde{h}\in\Pa$, the unique choice of $\delta$ is the trivial one.

\section{Remarks}

As already indicated above, we now discuss the conditions on the product (\ref{Md17}), induced on $\Pa^*$ by the coupling (\ref{Md16}), to be associative, that is:
\beq
\muts^*(f_1\otimes\muts^*(f_2\otimes f_3))(h)=\muts^*(\muts^*(f_1\otimes f_2)\otimes f_3)(h)
\eeq
with $f_i\in\Pa_{\theta}^*$ and $h\in\Pa_{\theta}$. Using (\ref{Md17}) we can compute both sides getting:
\bea
{\rm L.H.S.}=\Big(\muts^*(f_2\otimes f_3)\otimes f_1\Big)(\sigma\circ\Delta_{\theta}(h))=(f_3\otimes f_2\otimes f_1)\Big((\I\otimes\sigma)(\I\otimes\Delta_{\theta})(\sigma\circ\Delta_{\theta}(h))\Big)\\
{\rm R.H.S.}=\Big(f_3\otimes\muts^*(f_1\otimes f_2)\Big)(\sigma\circ\Delta_{\theta}(h))=(f_3\otimes f_2\otimes f_1)\Big((\sigma\otimes\I)(\Delta_{\theta}\otimes\I)(\sigma\circ\Delta_{\theta}(h))\Big)
\eea

Using compatibility of the coproduct with the multiplication map, $\Delta_{\theta}(h_1\cdot h_2)=\Delta_{\theta}(h_1)\cdot\Delta_{\theta}(h_2)$, and  co-associativity of the deformed coproduct, $(\I\otimes\Delta_{\theta})(\Delta_{\theta})(h)=(\Delta_{\theta}\otimes\I)(\Delta_{\theta}(h))$ we find that the associativity of the product $\muts^*$ translates into
\beq\label{Md24}
(\I\otimes\sigma)(\I\otimes\Delta_{\theta})(\sigma)=(\sigma\otimes\I)(\Delta_{\theta}\otimes\I)(\sigma)
\eeq

In the Hopf algebra cohomology, the condition above corresponds exactly to $\sigma$ being a {\it 2-cocycle} \cite{majid}. We have therefore shown, as we claimed, that the associativity of the product induced by the new pairing (\ref{Md16}) restricts in a cohomological way the possible choices of the map $\sigma$. In the case under discussion it is not necessary to check that the choice made in (\ref{Md22}) fulfills the condition given above since the induced product on spacetime is Moyal which is known to be associative. 

Another important point we can comment on here is the following. If we relax the condition of the multiplication is compatible with the left action of the Lorentz group, the whole family of multiplication maps (\ref{Md27}) is allowed. This allows us to get different deformations $\mathcal{A}_{\theta}(\mathbb{R}^4)$ of the algebras of functions on the spacetime choosing different $\delta$.

Now deformations of $\mathscr{F}(\mathbb{R}^4)$ are classified by deformation quantization cohomology (see \cite{Dito}) which is Hochshild cohomology. Also it can be shown \cite{Marmo} that maps $\delta$ in the same Hopf-algebra cohomology class (see \cite{majid}) induce isomorphic deformations. We thus get an action of Hopf algebra cohomology on deformation quantization cohomology. This action merits further study.

We want also to stress how the freedom in defining a map from $(H\otimes H)^*$ to $H^*\otimes H^*$, once the pairing $\langle\cdot,\cdot\rangle$ is given, is the only one available to characterize $H^*$ as a Hopf algebra. Both the product and the coproduct structures are then in fact canonically defined. The compatibility between the two, which characterizes completely $H^*$ as a Hopf algebra, can be also written in term of the pairings, although this is a very strong constraint\footnote{We gratefully acknowledge Alberto Ibort for having pointed this out.}. It can be shown, in fact, that the only product $\muts^*$ compatible with $\Delta_{\theta}^*$ (the coproduct induced on $\mathscr{F}(\Pa_{\theta})$ by the canonical pairing (\ref{Md9})) is the one given by the trivial choice $\sigma=\I\otimes\I$. However we are not interested in $\mathscr{F}(\Pa_{\theta})$ as a Hopf algebra. We are interested in it as an algebra and in recovering the Moyal plane therefrom. We are hence allowed to choose a non-trivial $\sigma$.

Finally we point out that the above construction can be generalized to any group $\mathscr{G}$ which is the semi-direct product of a subgroup $\mathscr{S}$ and translations.

\section{Conclusions}

The construction explained in this paper shows how the deformation of spacetime algebra is tightly connected with the deformation of the coproduct of its associated Hopf algebra (such as the Poincar\'e-Hopf algebra) . So far the latter was primarily seen as a consequence of the former one. We have now shown how the connection goes in the other direction as well. 

The deformations of Hopf agebras are classified by the Hopf algebra cohomology \cite{chari,majid}. The deformation of spacetime algebra is classified by Hochshild cohomology \cite{Dito}. The requirement that a deformed Hopf algebra acts on a deformed spacetime algebra when such an action exists in the undeformed case clearly puts constraints  on the possible deformations of both algebras and consequently ties the two cohomologies. It is of considerable interest to properly understand these connections. We have taken the first step in this direction by showing the two-way connection between the Moyal deformation of spacetime and the Drinfel'd twist of the Hopf-Poincar\'e algebra. We plan to more fully study these mutual dependences in later works.

\section{Acknowledgements}
We gratefully acknowledge discussions with Alexander Pinzul and Amilcar R. Queiroz. We also thank Anosh Joseph and Pramod Padmanabham for encouraging discussions. It is also a pleasure to thank Alvaro Ferraz and the Centro Internacional de F\'isica da Mat\'eria Condensada of Brasilia where this work was started and Alberto Ibort and the Universidad Carlos III de Madrid for their wonderful hospitality and support.

A.P.B. thanks T. R. Govindarajan and the Institute of Mathematical Sciences, Chennai for very friendly hospitality as well.

The work was supported in part by DOE under the grant number DE-FG02-85ER40231 and by the Department of Science and Technology, India.

\bibliographystyle{apsrmp}

\end{document}